%% file: PROC-CTD2020-52_arXiv.tex
\documentclass[10pt, paper=a4, UKenglish]{article}
\usepackage{graphicx}
\def\Title#1{\begin{center} {\Large #1 } \end{center}}
\def\Author#1{\begin{center}{ \sc #1} \end{center}}
\def\Address#1{\begin{center}{ \it #1} \end{center}}

\newcommand\pubblock{\rightline{\begin{tabular}{l} Proceedings of CTD 2020\\ \pubnumber\\
         \pubdate  \end{tabular}}}

\newenvironment{Abstract}{\begin{quotation} \begin{center}
             \large ABSTRACT \end{center}\bigskip
      \begin{center}\begin{large}}{\end{large}\end{center} \end{quotation}}

\newenvironment{Presented}{\begin{quotation} \begin{center}
             PRESENTED AT\end{center}\bigskip
      \begin{center}\begin{large}}{\end{large}\end{center} \end{quotation}}

\def\Acknowledgements{\bigskip  \bigskip \begin{center} \begin{large}
      \bf ACKNOWLEDGEMENTS \end{large}\end{center}}

\input econfmacros.tex

\usepackage{color}
\usepackage{lineno}
\usepackage{subfig}
\usepackage{hyperref}

\newcommand\pubnumber{PROC-CTD2020-52}

\newcommand\pubdate{\today}

\def\affiliation{
On behalf of the LHCb Real Time Analysis project,\\[2mm]
${}^{1}$ University of Cincinnati, Cincinnati, OH, United States\newline
${}^{2}$ Massachusetts Institute of Technology, Cambridge, MA, United States\newline
${}^{3}$ Princeton University, Princeton, NJ, United States}

\newcommand{\conference}{Connecting the Dots Workshop (CTD 2020)\\
April 20-30, 2020}

\usepackage{fancyhdr}
\definecolor{mygrey}{RGB}{105,105,105}
\begin{document}


\large
\begin{titlepage}
\pubblock

\vfill
\Title{An updated hybrid deep learning algorithm for identifying and locating primary vertices}
\vfill

\Author{Simon~Akar$^{1}$, Thomas~J.~Boettcher$^{2}$, Sarah~Carl$^{1}$, Henry~F.~Schreiner$^{3}$, Michael~D.~Sokoloff$^{1}$,
Marian~Stahl$^{1}$, Constantin~Weisser$^{2}$, Mike~Williams$^{2}$}
\Address{\affiliation}
\vfill

\begin{Abstract}
We present an improved hybrid algorithm for vertexing, that combines deep learning with conventional methods.
Even though the algorithm is a generic approach to vertex finding, we focus here on it's application as an alternative Primary Vertex (PV) finding tool for the LHCb experiment.

In the transition to Run\,3 in 2021, LHCb will undergo a major luminosity upgrade, going from 1.1 to 5.6 expected visible PVs per event, and it will adopt a purely software trigger.
We use a custom kernel to transform the sparse 3D space of hits and tracks into a dense 1D dataset, and then apply Deep Learning techniques to find
PV locations using proxy distributions to encode the truth in training data.
Last year we reported that training networks on our kernels using several Convolutional Neural Network layers yielded better than 90\,\% efficiency with no more than
0.2 False Positives (FPs) per event.
Modifying several elements of the algorithm, we now achieve better than 94\,\% efficiency with a significantly lower FP rate.
Where our studies to date have been made using toy Monte Carlo (MC), we began to study KDEs produced from complete LHCb Run\,3 MC data,
including full tracking in the vertex locator rather than proto-tracking.
\end{Abstract}

\vfill

\begin{Presented}
\conference
\end{Presented}
\vfill
\end{titlepage}
\def\thefootnote{\fnsymbol{footnote}}
\setcounter{footnote}{0}
%

\normalsize


\section{Introduction}
\label{intro}
The LHCb experiment is currently being upgraded for the planned start of Run\,3 of the LHC in 2021 to
facilitate recording proton-proton collision data at $\sqrt{s}=14\,$TeV with an instantaneous luminosity of $2\times10^{33}$\,cm$^{-2}$s$^{-1}$~\cite{LHCbCollaboration:2014vzo}.
Building on the success of Run\,2~\cite{Aaij:2018jht}, the experiment will continue moving towards a real-time-analysis approach and consequently remove the Level 0 hardware trigger in favor of a pure software trigger system.
In this process the the entire software stack is refactored, not only to reflect the new hardware components, but also to improve performance of reconstruction and selection.
One of the algorithms that is largely affected by the increased instantaneous luminosity is Primary Vertex (PV) finding.
The increased luminosity translates to an expected average number of PVs from 1.1 in Run\,2 to 5.6 in Run\,3, which will degrade the overall performance of PV reconstruction.
A new fast PV finding algorithm has been brought forward and is currently proposed as the baseline solution~\cite{Reiss20}.

\begin{figure}[!htb]
  \centering
  \includegraphics[width=0.7\linewidth]{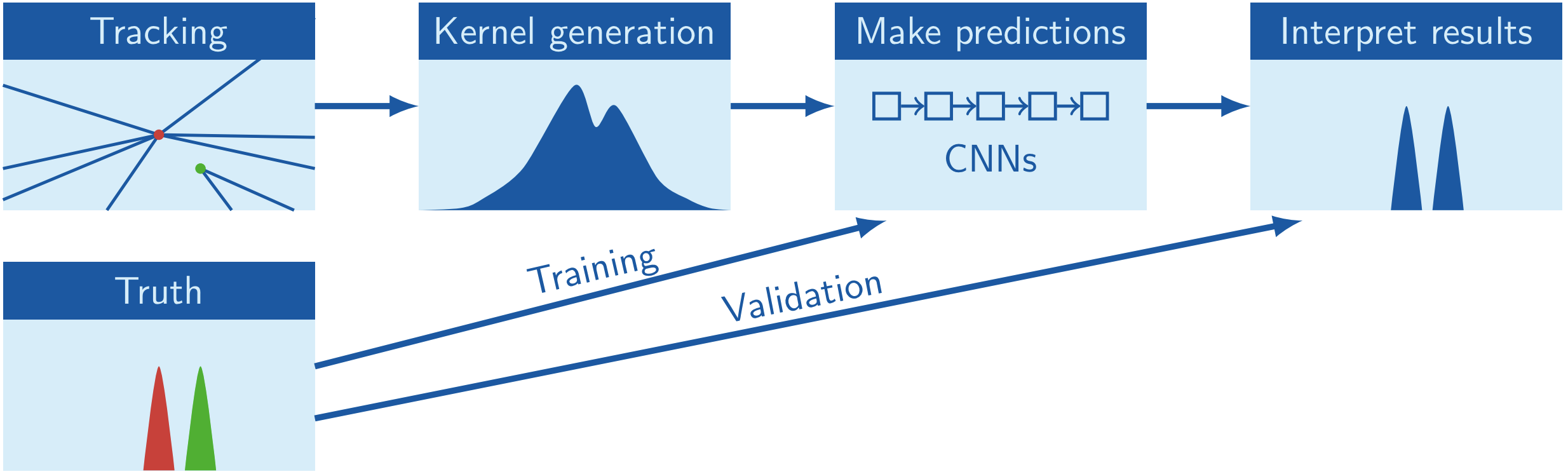}
  \caption{Schematic workflow of the hybrid deep learning algorithm for vertex finding.}
  \label{fig:Fig1}
\end{figure}

As an alternative we propose a hybrid algorithm, established in Ref.~\cite{Fang:2019wsd}, available on gitlab~\cite{repository}, and schematically shown in Fig.~\ref{fig:Fig1},
that approaches this challenge with machine learning techniques in the cluster search of PV finding - the part of the algorithm that drives the PV reconstruction efficiency.
In the context of PV finding at LHCb, the hybrid algorithm starts with tracks that have been reconstructed from hits in the Vertex locator (Velo).
Their location, direction, and covariance matrix information is used to reduce the problem from three to one dimension,
by calculating ``kernels'' in bins along $z$, \textit{i.e.} the beam direction.
The $z$ positions of the primary vertices are closely related to peaks in this kernel histogram, and it is then used as input to a Convolutional Neural Network (CNN)
to carry out the cluster search and predict the PV locations.
The output probabilities from the CNN are converted to a list of PV candidate $z$-locations using a simple peak finding algorithm.
A toy simulation of the LHCb detector is used to train, develop and test the algorithm.
In parallel, the algorithm has been deployed into the LHCb software stack where it can be used for PV finding.

\section{Kernel generation}
\label{sec:kg}
The kernel generation step converts sparse 3D data into a feature-rich 1D histogram. It consists of 4000 bins along the $z$-direction (beamline), each 100\,$\mu$m wide, covering the active area of the Velo around the interaction point. Each $z$ bin of the histogram is filled by the maximum kernel value in $x$ and $y$, where the kernel is defined by
\begin{equation}
\mathcal{K}(x,y,z) = \frac{\sum_\mathrm{tracks}\mathcal{G}(\mathrm{IP}(x,y)|z)^2}{\sum_\mathrm{tracks}\mathcal{G}(\mathrm{IP}(x,y)|z)} -
 \sum_\mathrm{tracks}\mathcal{G}(\mathrm{IP}(x,y)|z)\ .
 \label{eq:kernel}
\end{equation}
In Eq.~(\ref{eq:kernel}), $\mathcal{G}(\mathrm{IP}(x,y)|z)$ is a Gaussian function, centered at $x=y=0$ and evaluated at the impact parameter IP$(x,y)$:
the distance of closest approach of a track projection to a hypothesized vertex at position $x,y$ for a given $z$.
The width/covariance of $\mathcal{G}$ is given by the IP$(x,y)$ uncertainty/covariance matrix.
Finding the maximum $\mathcal{K}(x,y,z)$ is a two step process, where kernel values are first computed in a coarse $8\times 8$ grid in $x,y$;
then the parameters of that search are then taken as starting points for a \texttt{MINUIT} minimization process to find the maximum kernel value.

\begin{figure}[!htb]
  \centering
  \includegraphics[width=0.58\linewidth]{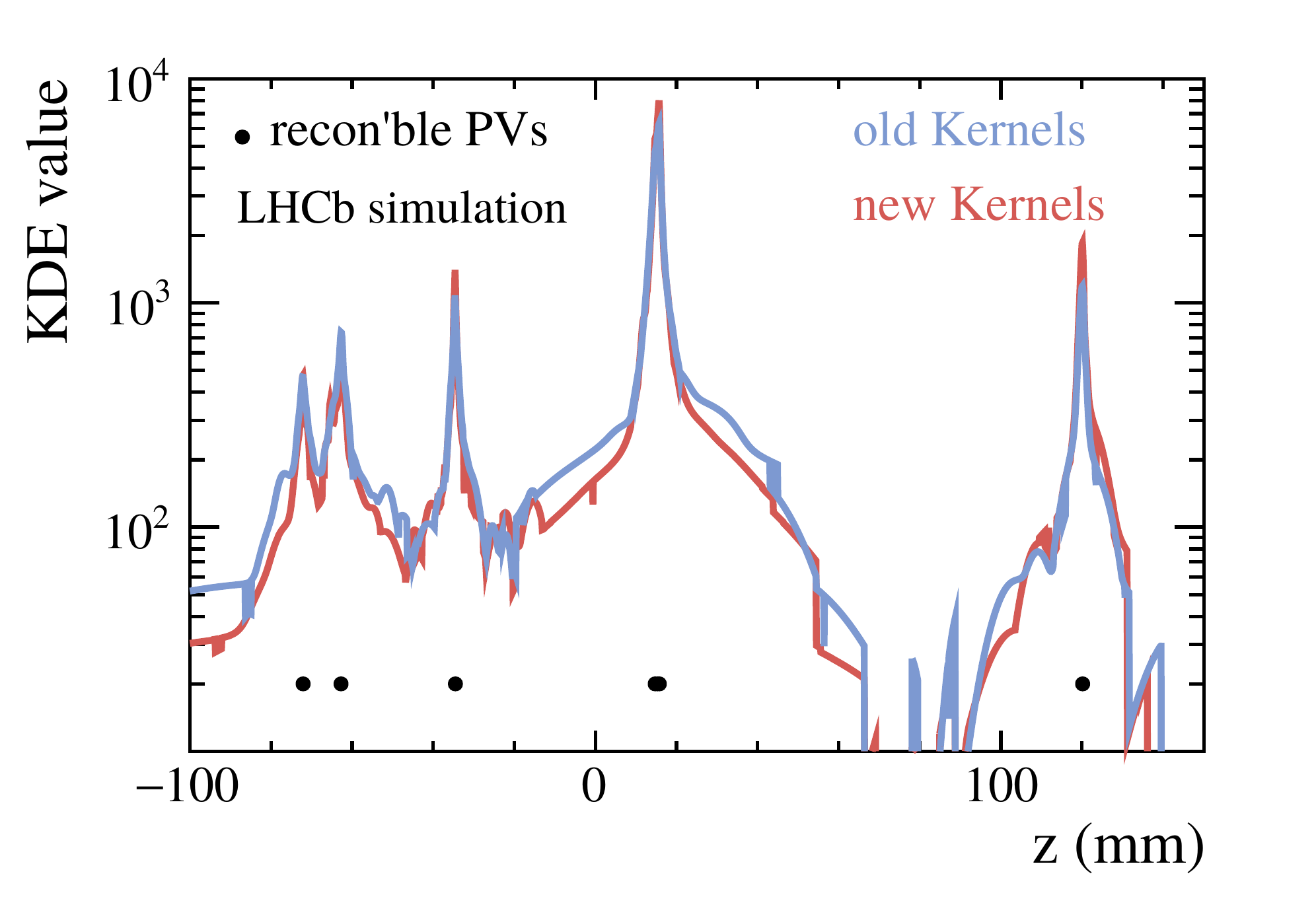}
  \caption{Comparison of kernel histograms heuristic (blue) and measured (red) IP$(x,y)$ uncertainty/covariance.
           A random single event with 6 reconstructible primary vertices is shown, whose $z$ positions are marked by black dots.
           Here, a simulated primary vertex is called reconstructible if more than four simulated tracks within the Velo acceptance emerge from that vertex.
           Note that there are 2 PVs at about $+15$\,mm, illustrating the issue of PV proximity that the cluster search faces frequently.}
  \label{fig:new_kernels}
\end{figure}

Equation~(\ref{eq:kernel}) is computed from the location, direction, and covariance matrix of input tracks.
This information can be provided by a standalone toy Monte Carlo proto-tracking~\cite{Fang:2019wsd} or from the proposed
LHCb Run\,3 production Velo tracking~\cite{Hennequin:2019itm}.
While the former uses a heuristic Gaussian width to estimate the IP$(x,y)$ uncertainty,
the latter makes use of the measured covariance matrix of the track state closest to beam to compute the IP$(x,y)$ covariance.
The difference between kernel histogram using the heuristic and measured IP$(x,y)$ uncertainty/covariance is exemplarily shown in Fig.~\ref{fig:new_kernels}.
It can be seen that the kernels are more pronounced around PVs, which suggests that -- once re-trained -- the CNN performance should increase.
Further improvement with LHCb simulation data is expected from centering $\mathcal{G}$ at the actual beamline position at a given $z$ position, and
by using Kalman-fitted Velo tracks as input to the kernel generation.

\section{CNN based cluster search}
\label{sec:cs}
The cluster search in our algorithm is conducted by a Convolutional Neural Network (CNN) consisting of several one dimensional convolution layers.
We use \texttt{PyTorch}~\cite{NEURIPS2019_9015} for training and development, and \texttt{TorchScript}~\cite{TorchScript} as \texttt{C++} inference engine in
the LHCb software stack.

\begin{figure}[!htb]
  \centering
  \includegraphics[width=0.6\linewidth]{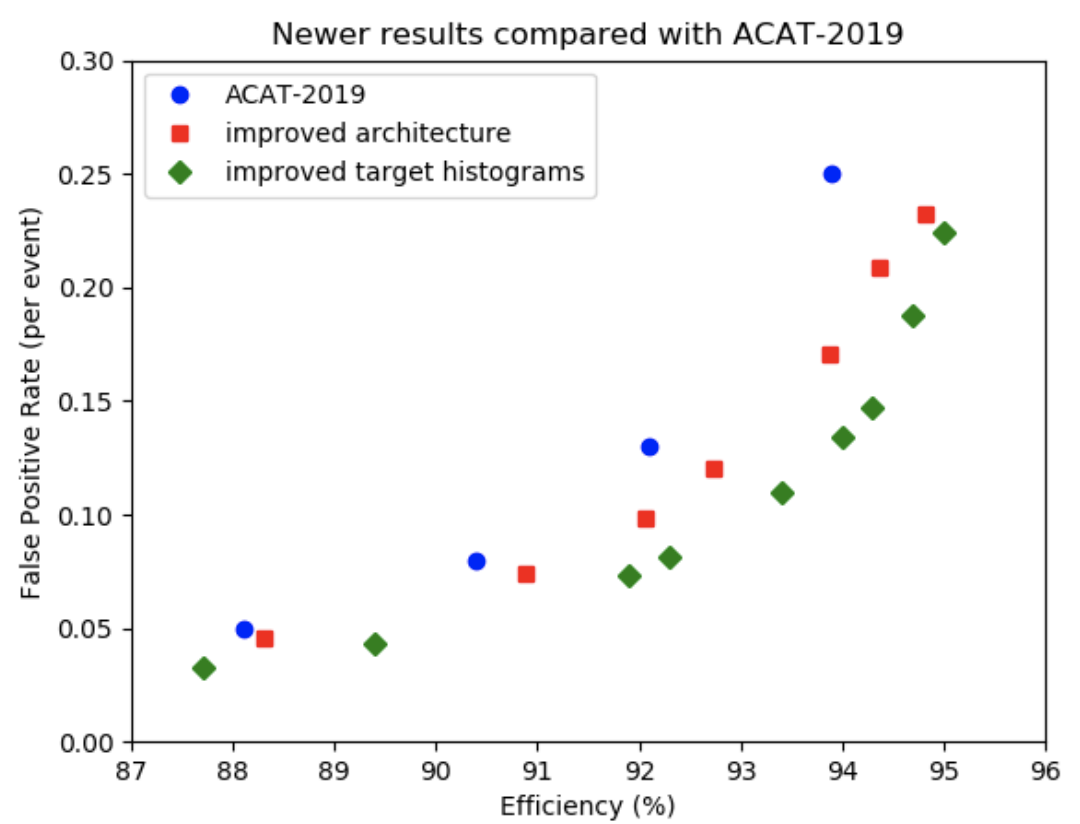}
  \caption{Performance improvements with respect to Ref.~\cite{Fang:2019wsd} achieved by modifying target histograms,
  adding layers to the CNN and adding $x$, $y$ position information perturbatively. Details are described in the text.}
  \label{fig:performance_improvements}
\end{figure}

Before being able to train a CNN, we need to define what it should learn, \textit{i.e.} give it a target.
The CNNs in our algorithm is designed to output a 4000 bin \textit{target histogram}, just as the input kernel histogram, where essentially all ``noise'' is removed.
A first approach to create such a target histogram is to define a Gaussian with unit area and a width of one bin (100\,$\mu$m) centered around the simulated PV position.
These simple target histograms have been updated to take the PV resolution as a function of track multiplicity into account, which affects width and area of the Gaussians.
The resulting performance improvement is plotted in Fig.~\ref{fig:performance_improvements}.
Further, regions around PVs that do not pass the criterion of being reconstructible, \textit{i.e.} have 5 or less detectable associated tracks, are ``masked''.
Masked regions are effectively hidden during training, so that discoveries in them are neither punished nor rewarded.

Our best performing network to date is composed of 6 convolutional layers with leaky ReLU activation functions in-between hidden layers and a
softmax activation for the output layer.
The widths and padding of each convolution kernel was chosen by visual inspection of the data; the number of channels were increased until benefits
were no longer noticeable upon adding new channels.
The training is carried out on GPUs using mini-batch gradient descent, the Adam optimizer and dropout regularization.

A custom cost function, similar to cross-entropy, has been defined and it was found that it's initial symmetric form favored small
false positive rates at the expense of efficiency. Therefore, a single parameter asymmetry term has been added to the cost function,
serving as powerful control for selecting the false positive to efficiency tradeoff~\cite{Fang:2019wsd}.
To stabilize the training process in early epochs, the last convolution layer is replaced by a Fully Connected (FC) layer.
After several training epochs with this architecture, the weights of the convolutional layers are fixed, the FC layer is replaced by a convolutional layer,
and the network is trained for a few more epochs. Then, all weights are floated and the CNN is trained in it's final architecture.

Recently, we improved the CNN performance by adding $x$ and $y$ position information, found by maximizing the kernel (Eq.~(\ref{eq:kernel})), in a perturbative manner.
Adding the information perturbatively is important because the CNN overestimates the importance of these variables compared to the $z$ information.
The perturbative addition works as follows: On the basis of the original network, another CNN with 3 convolutional layers that solely
processes $x$ and $y$ position information is trained independently, but parallel to the original network.
Such, that both CNN responses at the end of each training epoch are multiplied.
The CNN with $x$ and $y$ information will contribute with values close to 1 in most cases, but can veto kernel peaks with high $x,y$
gradients that have been observed in data and can lead to false positives.
The performance improvement of the perturbation network, together with the addition of one convolutional layer is shown in Fig.~\ref{fig:performance_improvements}.

\section{Results}
The proof of principle, that the here proposed hybrid deep learning algorithm is an efficient vertex finding tool has been established in Ref.~\cite{Fang:2019wsd}.
Since then, the algorithm's performance has been improved further by changes described in Sec.~\ref{sec:cs} and are shown in Fig.~\ref{fig:performance_improvements}.
We want to highlight that for a fixed PV finding efficiency of 94\,\%, the false positive rate per event could be reduced by a factor of 2.
These numbers have been obtained on toy simulation using metrics that do not fully reflect standard LHCb definitions,
and are thus not directly comparable to the ones given in Ref.~\cite{Reiss20}.
However, first studies using official LHCb simulation in place of the toy simulation with our metrics agree with those from pure toy simulation.
We are looking forward to re-train the CNNs with LHCb simulation data and expect further improvements of the PV finding performance.

\section{Conclusions and Outlook}
We have presented a hybrid deep learning vertexing algorithm and it's application as PV finding algorithm in the LHCb Run\,3 software trigger.
The algorithm has been privately deployed in the LHCb CPU software stack and it's performance has been further improved using toy data.

Future milestones of the project are well defined.
We plan to deploy the algorithm in Allen, the High‑Level Trigger application on GPUs for LHCb~\cite{Allen}.
Further, it is of great importance to have a one to one comparison with the currently proposed LHCb Run\,3 production PV finding algorithm~\cite{Reiss20}.
This means that we need to refactor the current implementation of our algorithm in the LHCb software stack to be able to run the same performance benchmark tests.
In parallel, we need to re-train our CNN with official LHCb simulation data in place of toy data.
We expect further performance improvements from doing so,
in particular from using the measured covariance matrix of (Kalman-fitted) Velo tracks in the generation of kernel histograms.
Currently, the generation of these kernel histograms is the throughput bottleneck of our algorithm due to the usage of \texttt{MINUIT}.
We are planning to replace the kernel generation by a fast machine learning algorithm,
which could be merged with the cluster search CNN into a single algorithm that predicts PV positions from the output of the Velo track reconstruction in one step.
We are also investigating pruning techniques to speed up the throughput of the inference.
Moreover, we plan to associate Velo tracks with found PV candidates probabilistically to reduce the false positive rate.
In an adjacent step, we plan to probabilistically identify secondary vertices and associated tracks.

\Acknowledgements
This work was supported by the National Science Foundation under Cooperative Agreement
OAC-1836650, OAC-1739772, and OAC-1740102. It was also supported by the University of
Cincinnati Women in Science and Engineering program.


\end{document}

%% file: econfmacros.tex
\def\beq{\begin{equation}}
\def\eeq#1{\label{#1}\end{equation}}
\def\eeqn{\end{equation}}

\def\beqa{\begin{eqnarray}}
\def\eeqa#1{\label{#1}\end{eqnarray}}
\def\eeqan{\end{eqnarray}}

\let\bar=\overbar

\def\Dslash{\not{\hbox{\kern-4pt $D$}}}
\def\dslash{\not{\hbox{\kern-2pt $\del$}}}

\def\msb{{\bar{\ssstyle M \kern -1pt S}}}




%% file: PROC-CTD2020-52_arXiv.bbl
\begin{thebibliography}{99}


\bibitem{LHCbCollaboration:2014vzo}
LHCb Collaboration,
"LHCb Trigger and Online Upgrade Technical Design Report,''
\href{https://cds.cern.ch/record/1701361}{CERN-LHCC-2014-016, LHCb-TDR-2014-016}.

\bibitem{Aaij:2018jht}
R.~Aaij \textit{et al.} [LHCb],
"Design and performance of the LHCb trigger and full real-time reconstruction in Run 2 of the LHC,''
\href{doi:10.1088/1748-0221/14/04/P04013}{JINST \textbf{14}, no.04, P04013 (2019)};
\href{https://arxiv.org/abs/1812.10790}{[arXiv:1812.10790 [hep-ex]]}.

\bibitem{Reiss20}
F.~Rei\ss, "Fast parallel Primary Vertex reconstruction for the LHCb Upgrade,'' \href{https://indico.cern.ch/event/831165/contributions/3717129/}{Talk given at this conference};
\href{https://cds.cern.ch/record/2717609}{LHCb-TALK-2020-044}. Publication in preparation.

\bibitem{Fang:2019wsd}
R.~Fang, H.~F.~Schreiner, M.~D.~Sokoloff, C.~Weisser and M.~Williams,
"A hybrid deep learning approach to vertexing,''
\href{https://arxiv.org/abs/1906.08306}{[arXiv:1906.08306 [physics.ins-det]]}.

\bibitem{repository}
\texttt{pv-finder} repository: \url{https://gitlab.cern.ch/LHCb-Reco-Dev/pv-finder}.

\bibitem{Hennequin:2019itm}
A.~Hennequin, B.~Couturier, V.~Gligorov, S.~Ponce, R.~Quagliani and L.~Lacassagne,
"A fast and efficient SIMD track reconstruction algorithm for the LHCb Upgrade 1 VELO-PIX detector,''
\href{https://arxiv.org/abs/1912.09901}{[arXiv:1912.09901 [physics.ins-det]]}.

\bibitem{NEURIPS2019_9015}
A. Paszke \textit{et al.} [The PyTorch team], "PyTorch: An Imperative Style, High-Performance Deep Learning Library,''
\href{http://papers.neurips.cc/paper/9015-pytorch-an-imperative-style-high-performance-deep-learning-library.pdf}{Advances in Neural Information Processing Systems 32, 8024 (2019).}

\bibitem{TorchScript}
The PyTorch team, "Torch Script,'' \url{https://pytorch.org/docs/stable/jit.html}.

\bibitem{Allen}
R.~Aaij \textit{et al.}, "Allen: A High‑Level Trigger on GPUs for LHCb,''
\href{doi:10.1007/s41781-020-00039-7}{Comput. Softw. Big Sci. \textbf{4}, no.1, 7 (2020)};
\href{https://arxiv.org/abs/1912.09161}{[arXiv:1912.09161 [physics.ins-det]]}.

\end{thebibliography}
